\documentclass[a4paper,fleqn]{cas-sc}
\usepackage[numbers]{natbib}

\usepackage{color}
\usepackage{graphicx}

\usepackage[dvipsnames]{xcolor}
\usepackage{cancel}

\usepackage{amsmath,amssymb}

\begin{document}
\title[mode=title]
  {Spinodal Instability at the Onset of Collective Expansion in Nuclear Collisions}

\author[1]{Pawel Danielewicz}[orcid=0000-0002-1989-5241]
\ead{danielewicz@nscl.msu.edu}

\author[1]{Hao Lin}

\author[2,3]{Jirina R.~Stone}

\author[4]{Yoritaka Iwata}

\address[1]{National Superconducting Cyclotron Laboratory and
Department of Physics and Astronomy, Michigan State University, East Lansing, Michigan 48824, USA}
\address[2]{Department of Physics (Astrophysics), University of Oxford, Oxford OX1 3RH, United Kingdom}
\address[3]{Department of Physics and Astronomy, University of Tennessee, Knoxville, TN 37996, USA}
\address[4]{Faculty of Chemistry, Materials and Bioengineering, Kansai University, Osaka, 564-8680, Japan}

\begin{abstract}[S U M M A R Y]
Using transport theory to model central Au + Au collisions in the energy region of 20 - 110 MeV/u, at impact parameters $b\leq$ 5 fm, we predict a measurable impact of the spinodal instability as the collective expansion sets in with energy.  Two transport models are employed, the pBUU model, solving a Boltzmann-Uehling-Uhlenbeck equation,  and the Brownian Motion (BM) model, solving a set of Langevin equations to describe the motion of individual nucleons in a noisy nuclear medium. We find without ambiguity, for the first time, that a combination of delayed equilibration, onset of collective expansion and the spinodal instability produces a pair of transient ring structures, made of the projectile and target remnants, with spectator nucleons predicted to end in the entities reminiscent of stones in jewelry, on the rings. The ring structures, calculated in the configuration space and mapped onto the velocity space, could be detected in experimental collective flow data.
\end{abstract}
\maketitle 

During the last decades, the thermodynamics of matter created in heavy-ion (HI) collisions has been studied in a variety of theoretical approaches, see e.g.~\cite{chomaz_nuclear_2004,rios_liquid-gas_2008,mukherjee_variational_2009,taranto_selecting_2013}, and in numerous experiments, e.g.~\cite{reisdorf_collective_1997,reisdorf_dynamics_2000,andronic_excitation_2005}, at a wide range of beam energies, impact parameters and projectile-target combinations.  In drawing conclusions from the data for matter densities in excess of the normal density $\rho_0=0.16 \, \text{fm}^{-3}$, the ability to model the collisions in transport theory has been critical for relating the Equation of State (EOS), and other bulk properties of nuclear matter, to measurable signals coming from the collisions.

Collision dynamics and the region of EOS that can be explored is regulated by the beam energy and, to some degree, by the impact parameter \cite{danielewicz_effects_1995}.  Heavier systems generally allow for more equilibration and thus more direct EOS relevance.  Mass-symmetric systems facilitate dealing with any center of mass effects in the experiment.  At low energies, but high compared to Coulomb barrier, the compression of matter is modest.  After density relaxes, the surface tension can deliver enough pull for the projectile and target remnants to deflect, on an average basis, in such a manner as for attractive interactions in two-body scattering, to negative scattering angles.  Overall emission tends to be focused around the reaction plane, reminding emission from a rotating compound nucleus formed in central region when nuclei fuse close to the Coulomb barrier.  As central density and entropy grow with energy, so do the transverse components of the pressure tensor.  From the incident energies of 50--100$\, \text{MeV/u}$ on \cite{magestro_disappearance_2000,andronic_excitation_2005,borderie_radial_1996}, these pressure components exert enough push on particles departing the system to leave clear specific traces of transverse collective motion in transverse emission spectra.  The average deflection in the reaction plane starts to point towards positive scattering angles, as for repulsive interactions in two-body scattering.  As the motion becomes supersonic with increasing energy, progressively cleaner separation emerges between the matter, called participant, that faces some matter from the opposing nucleus, and the matter that faces no opposition, called spectator.  The spectator matter shadows early emission from the participant matter.  At low longitudinal velocity, emission in transverse directions becomes more intense out of the reaction plane than in the plane.

\begin{figure*}
\centerline{\includegraphics[width=.5\linewidth]{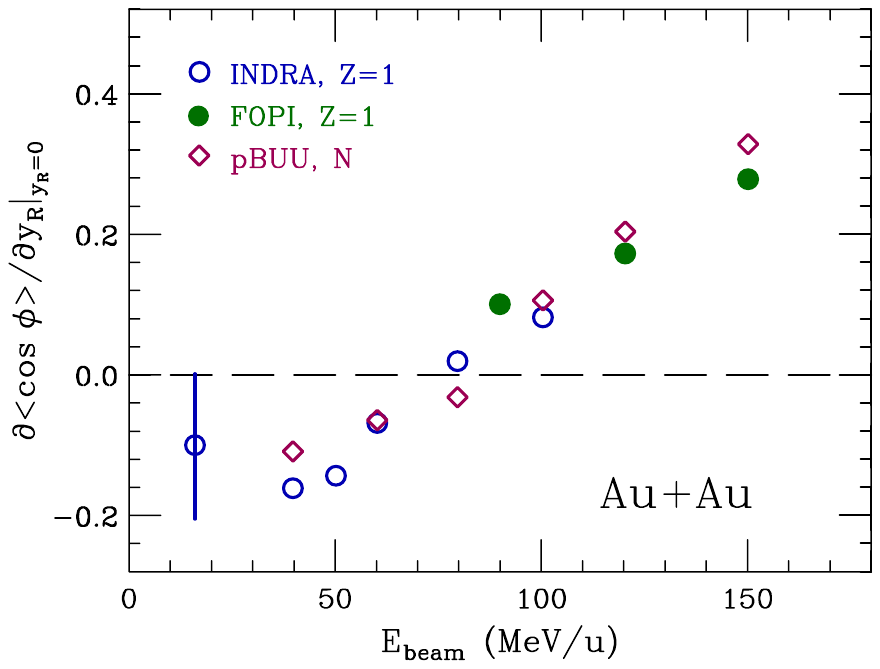}}
\caption{Excitation function of the sideward flow in semi-central Au + Au collisions. The flow is quantified by derivative of the first-order azimuthal Fourier coefficient with respect to the center-of-mass rapidity normalized to the beam, $y_R$.  The circles display measurements for the $Z=1$ particles with the impact parameter $b=$(2--5.5$) \, \text{fm}$ collisions by the INDRA (open) \cite{andronic_systematics_2006} and FOPI (filled) \cite{PhysRevC.64.041604} Collaborations. The diamonds show results for nucleons from pBUU calculations at the representative $b=4 \, \text{fm}$.}
\label{fig:Slopes}
\end{figure*}

Experimentally, collective motion is quantified relative to the reaction plane for the system and anisotropies in that motion are exploited to tell the reaction plane direction.  The component of the collective motion which is isotropic in the azimuthal angle $\phi$, relative to the reaction plane, is termed radial flow.  The anisotropies in $\phi$ are quantified in terms of Fourier coefficients for particles, such as $\langle \cos{\phi} \rangle$, or powers of momentum components, such as $\langle p^x \rangle = \langle p^\perp \cos{\phi} \rangle$.  The evolution of deflection with incident energy, or the sideward flow, can be, for example, investigated using the derivative of the first coefficient $\text{d}\langle \cos{\phi} \rangle / \text{d}  y_R  |_{y_R=0}$, where $y_R = y/y_\text{Beam}$ is the center of mass rapidity normalized to the beam.  The midrapidity focusing around the reaction plane, or the elliptic flow, can be quantified by the second coefficient $\langle \cos{2 \phi} \rangle_{y_R=0}$.  Fig.~\ref{fig:Slopes} shows, in particular, the evolution of the sideward flow for the Au + Au system, exhibiting the change in sign of the deflection as pressure builds up in the energy region of interest in this paper.  Notably, in the experiment~\cite{LUKASIK2005223,PhysRevC.64.041604} only magnitude of the sideward flow is determined, and the change in sign is inferred from the vanishing of the flow \cite{magestro_disappearance_2000}.  More information on systematics of collective flow in the heavy Au + Au system can be found in~\cite{LUKASIK2005223,PhysRevC.64.041604,borderie_radial_1996}.  Systematics of the anisotropic flows at high incident energies, of few hundred MeV/u or more, have been used to constrain EOS at supranormal densities~\cite{Danielewicz:2002pu,le_fevre_constraining_2016}.  The efforts to learn about the EOS, particularly the pressure as a function of density, have been complicated both by the involved collision physics, not yet fully comprehended,  and by the unavoidable simplifications in transport tools used in their description. The~limited duration of the collisions is likely to lead to an incomplete equilibration and finite system sizes require careful treatment of surface effects.

Yields of different species, including nucleons, light clusters (mass number $A \le 4$) and intermediate mass fragments (IMF; charge number $Z \gtrsim 3$) from the reaction also contain important information about the dynamics and potentially the EOS. Increased emission of IMF is usually seen as a fingerprint of a possible liquid-gas phase transition, predicted to occur in equilibrated nuclear matter at subnormal densities~\cite{danielewicz_shock_1979,rios_liquid-gas_2008}.  However, definitive experimental evidence of the transition has been lacking~\cite{reisdorf_dynamics_2000,indra_and_aladin_collaborations_bimodal_2009,le_fevre_bimodality:_2008} because the increased production of IMF could be also explained by mechanisms not invoking the phase transition, including mechanical fractures, sequential compound system decays and coalescence.

\begin{figure*}
\centerline{\includegraphics[width=.5\linewidth]{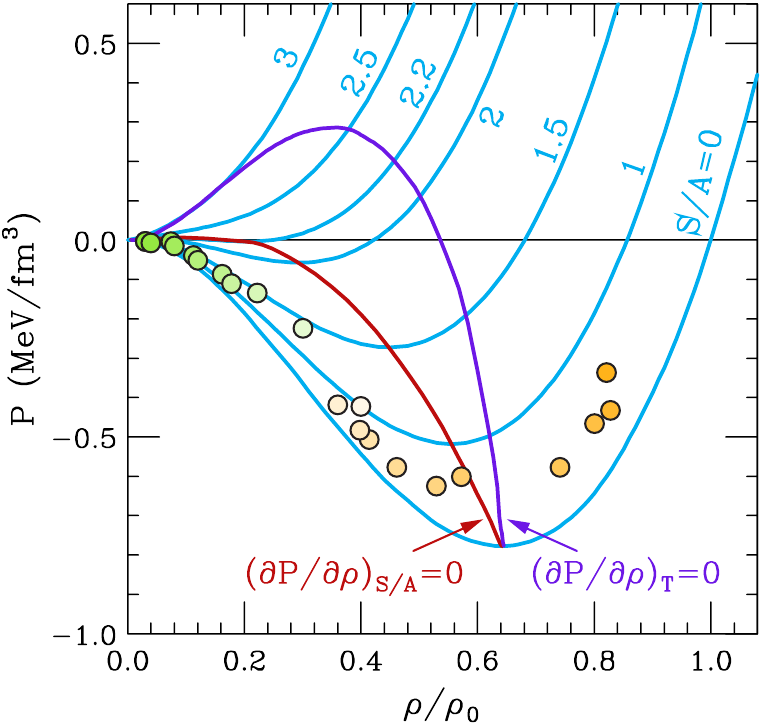}}
\caption{Adiabats for equilibrated symmetric nuclear matter, in the pressure-density plane, $P$-$\rho/\rho_0$, represented by the lines marked with entropy per nucleon values, $S/A$.  Boundaries of the low-entropy regions of adiabatic and isothermal instabilities, $(\partial P/\partial \rho)_{S/A}=0$ and $(\partial P/\partial \rho)_T=0$, respectively, are shown, too.  The matter is described here with the energy functional of the pBUU transport model~\cite{danielewicz_determination_2000}.  The circles illustrate matter in the vicinity of two representative nucleons from the pBUU simulations of Au + Au collisions at $60 \, \text{MeV/u}$ and $b=2 \, \text{fm}$.  One of those nucleons ends up in the middle of the stone of a late-stage ring forming in the collision; the other ends up near the overall center of the ring.  The circles, that illustrate matter for the two respective nucleons, cluster either towards the higher or lower densities in the figure and they represent the times from $100$ to $200 \, \text{fm}/c$, at $10 \, \text{fm}/c$ increments, with the fill intensifying with time (yellow and green fill, respectively, in the color version of the figure).  The pressure values for the surrounding matter stem from the trace of local momentum-flux tensor.}
\label{fig:Prho}
\end{figure*}

One possibility of an unambiguous demonstration of the transition results from examining the effect of the spinodal volume instability in the transition region~\cite{chomaz_nuclear_2004,sasaki_density_2007}.  Once matter enters the unstable region, non-uniformities in the order parameter for the transition, such as density, are rapidly amplified until stable coexisting phases are reached. Such a scenario is applicable in a wide range of physical phenomena, for example binary fluids or solids or hadronization of quark-gluon plasma.  The region of the spinodal instability in density $\rho$, of interest for the HI collisions, is indicated in Fig.~\ref{fig:Prho}, for equilibrated neutron-proton symmetric matter, as that of negative adiabatic compressibility, $(\partial P/\partial \rho)_{S/A}<0$.  Here, $P$ is pressure and $S/A$ is entropy per nucleon.  The adiabatic compressibility matters, as passage of matter through low-density region in collisions is rapid and largely adiabatic.  Amplified density differences for a system diving into that region can lead to fragmentation and the proof of the instability playing the role could lie in tying the pattern of fragmentation to the dynamics.  With proximity to axial symmetry in the geometry of near-central collisions, geometric structures in fragmentation can include cigar shapes, disks and rings.  For reference, we further indicate in Fig.~\ref{fig:Prho} the region of negative isothermal compressibility, which ends at the anticipated critical point for the nuclear liquid-gas phase transition \cite{danielewicz_shock_1979}.

Surface Rayleigh instabilities \cite{john_william_rayleigh_scientific_1964} can compete with volume instabilities in a patterned fragment formation.  However, the surface instabilities typically operate when systems reach densities close to normal, with the characteristic situation being fission, while the volume instabilities exclusively operate at low densities, $\rho \lesssim 2 \rho_0/3$, cf.~Fig.~\ref{fig:Prho}.  This underscores the importance of inspecting density magnitude when deciding on the instability type.  When both types of instability are there, the unstable volume modes grow faster \cite{napolitani_dynamic_2019}.

Early transport models~\cite{moretto_new_1992,bauer_bubble_1992,danielewicz_effects_1995,norbeck_nuclear_1996} predicted that, in head-on collisions, the nuclei would stop on each other and form a thin disk perpendicular to the beam axis.  Interaction between the surfaces on the two sides of the thin disk and an associated Rayleigh instability was invoked to explain the disk evolving into a ring that further fragmented into drops due to surface instabilities~\cite{moretto_new_1992}.  Volume instability was invoked too \cite{bauer_bubble_1992,li_dynamical_1993,norbeck_nuclear_1996}, but rather in the context of drop formation in the middle of the disk structure, than the ring formation.  Eventually, systematic experimental studies of the stopping in energetic collisions of heavy nuclei revealed that the nuclei fail to completely stop even in the most central collision~\cite{andronic_systematics_2006,zhang_comparison_2018}, i.e., the step preceding disk and ring formation in the simulations is not happening.

Admittedly, theoretical challenges remain in transport models. Realistic quantum theory for the collisions is beyond the current means.  Semi-classical transport models necessarily resort to compromises and choices, depending on the physics questions to be addressed. To adequately represent the collisions, where matter is generally out of equilibrium, the input to transport models has been adjusted to match experiments and an extrapolation to equilibrium for reaching conclusions has been used~\cite{Danielewicz:2002pu,le_fevre_constraining_2016}.  The recent comparison of transport codes against each other and the known constraints, strengthened their predictive power~\cite{xu_understanding_2016,zhang_comparison_2018}.
The transport codes moved to a new level of sophistication such as including collective oscillations around ground states~\cite{wang_nuclear_2019}.  Still, specific physics limitations remain. Models based on solving the semi-classical Boltzmann equation cannot describe production of IMF, and molecular dynamics models cannot describe production of light clusters.  Hybrid approaches are employed to overcome these shortcomings~\cite{tan_fragment_2001,le_fevre_constraining_2016}.  Hardly ever does the theory serve as a substitute for the experiment and the main role remains to provide guidance for the experiments to validate.

In this work we further explore the impact of a liquid-gas phase coexistence and spinodal instability in the collision (see also~\cite{napolitani_dynamic_2019}), looking for a unique experimental signal that would demonstrate the presence of these effects.  We study Au+Au collisions in two contemporary transport models, pBUU~\cite{danielewicz_production_1991,danielewicz_determination_2000,zhang_comparison_2018}, which solves the semiclassical Boltzmann-Uehling-Uhlenbeck equation, and the Brownian Motion (BM) model~\cite{lin_one-body_2019}, in which the beyond-mean-field dynamics in collision is reformulated in terms of a one-body  Brownian motion of nucleons in the nuclear medium, in contrast to the traditional two-body scattering. The major difference between the two models is that in pBUU correlations are perceived to relax quickly while in BM a possible lasting impact of the correlations is simulated with fluctuations.

\begin{figure*}
\includegraphics[width=\linewidth]{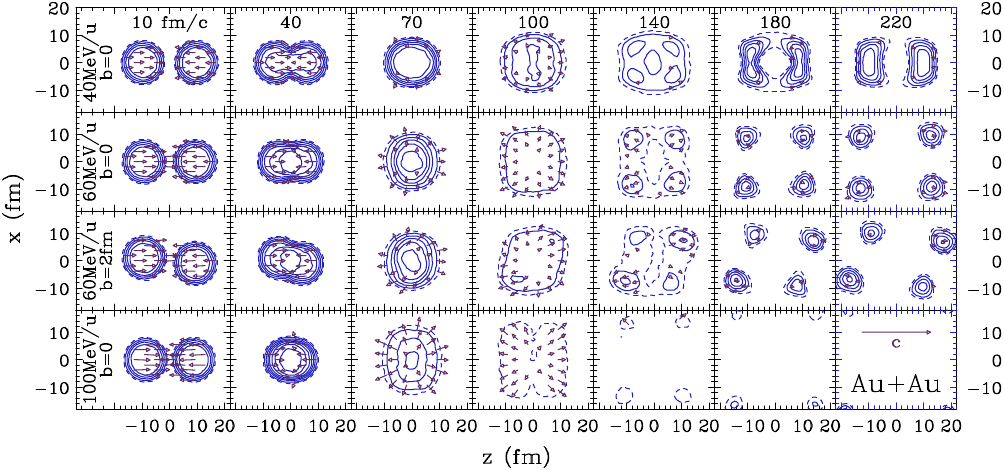}
\caption{Contour plots of the baryon density in the reaction (z,x) plane at different times in the center of mass of the Au + Au system colliding at different beam energies at two centralities, as predicted by the pBUU simulation. The $z$-axis is the collision (beam) axis.  The top, second and bottom rows represent results of a head-on collision at 40, 60, 100 MeV/u incident energy. The third row demonstrates the effect of a non-zero impact parameter at 60 MeV/u. The columns indicate different times. In each panel, the outer, dashed contour represents $0.1\rho_0$, and the subsequent solid contours are at the increments of $0.2\rho_0$, starting with $0.2\rho_0$.  The arrows illustrate collective velocity for the matter at selected locations.}
\label{fig:contour}
\end{figure*}

The beam energies in the simulation are selected to best test the expected spinodal conditions at low entropy ($S/A \sim 1$) when collective expansion sets on.
Fig.~\ref{fig:contour} shows contour plots of the baryon density $\rho$,  as predicted by pBUU, in the reaction plane at different times in the center of mass frame, during head-on Au + Au collisions at incident energies of 40, 60 and 100 MeV/u and $b= 2 \, \text{fm}$ collision at 60 MeV/u. The arrows illustrate the collective velocity field in the reaction plane. As nuclei overlap in the collision, the particle density increases above~$\rho_0$.  The matter becomes excited and expansion and eventual emission into the vacuum sets in. While the equilibration progresses and the energy of the relative motion turns into excitation, the nuclei fail to stop even in the head-on collisions, consistent with data~\cite{andronic_systematics_2006}, i.e., the equilibration is incomplete during the nuclear overlap.  As the collision progresses, the remnants of the original nuclei move with a much reduced velocity as compared to the original nuclei. These remnants are expanding and highly flattened following the impact against the opposing nucleus. The expansion slows down as the matter thins and cools down.

At times $t \sim 120 \, \text{fm}/c$, the maximal densities in the displayed head-on collisions are as low as $\rho_\text{max} \simeq 0.45, 0.33$, and $0.20 \, \rho_0$ at the beam energies of 40, 60 and $100 \, \text{MeV/u}$.  At such low densities, transverse components of internal pressure are negative, cf.~Fig.~\ref{fig:Prho}.  With negative pressure from the interior side and none outside, the front of the matter expanding transversally stalls and density there grows relative to the interior.  Since the matter is in the region of spinodal instability, the matter at somewhat higher density depletes the matter  at lower density, i.e., effectively the matter accumulates at the edges, so the remnants of the original nuclei turn into rings.  For reference, we use representative nucleons as tracers for matter, one ending up in the structure on a ring forming in a $b=2 \, \text{fm}$ $60 \, \text{MeV/u}$ collision, cf.~Fig.~\ref{fig:contour}, reminiscent of a decorative stone in jewelry, and another in the middle of the ring.  We trace back the state of the matter in the vicinity of those nucleons and represent that state in Fig.~\ref{fig:Prho}, from 100 to $200 \, \text{fm}/c$.  While the two nucleons start at similar density around $t=100 \, \text{fm}/c$, in a seemingly structureless lump of matter, they meet very different late-term fates, consistent with expectations on the instability.  We may note that interaction of surfaces is out of the question around that time in ring formation, though surface instabilities are likely to play a role later, after the rings become compact. A surface instability cannot explain a hollow formation inside matter ($t \sim 140 \, \text{fm}/c$).

Although there is a similarity with predictions of the early models~\cite{moretto_new_1992,bauer_bubble_1992,li_dynamical_1993,danielewicz_effects_1995,norbeck_nuclear_1996}, the new result is the prediction of formation of two disjoint rings in the central collisions, rather than one.  These rings form out of the matter that equilibrates late in the collision rather than early.  The matter is highly thinned and completely immersed in the region of spinodal instability in the $P$--$\rho$ diagram.  While ring formation was attributed in the past to a Rayleigh instability, there is no ambiguity here with the spinodal instability being behind ring formation.

Details and fate of the rings beyond $t \sim 120\, \text{MeV/u}$ depend on the incident energy and the centrality of the collision. In head-on ($b=0$) collisions, the rings are symmetric and perpendicular to the beam axis.  At low beam energies the expansion is slow enough so that negative pressure can stall it to a halt.  Surface tension can further pull the incipient ring structures in and gradually evolve the shape of the reaction products towards spherical.  At the highest beam energies (from $90 \, \text{MeV/u}$ on) the expansion is vigorous, the matter thins out quickly and the rings fail to form a pattern that might dominate over fluctuations expected in the final stages of collisions and missing so far from the simulations. At intermediate energies around 60 MeV/u and above, the expansion just gains strength.  It gets then, on the one hand, strong enough to prevent the rings from collapsing back onto a single compact shapes but, on the other hand, weak enough so that the nuclear density has time to grow over significantly large distances and values, making the rings persist rather than fleet.

At finite but low impact parameters ($b \lesssim 4 \, \text{fm}$), the rings continue to form at 60 MeV/u and above, but are tilted towards the beam axis and are asymmetric in shape. The matter that plays the spectator role in the supersonic context at much higher energies contributes to the thicker leading portions or `stones' of the rings.  If the impact parameter is increased further, the back sides of the rings continue thinning in the same fashion as the $b=0$ rings when the energy is increased above threshold for the ring appearance. This is important because it is easier to isolate a broader range of impact parameters experimentally - the rings with grossly thinned out backs in the pBUU calculations may realistically survive as croissant shapes in the final stages of the collision.

\begin{figure}
\centering
\includegraphics[width=\linewidth]{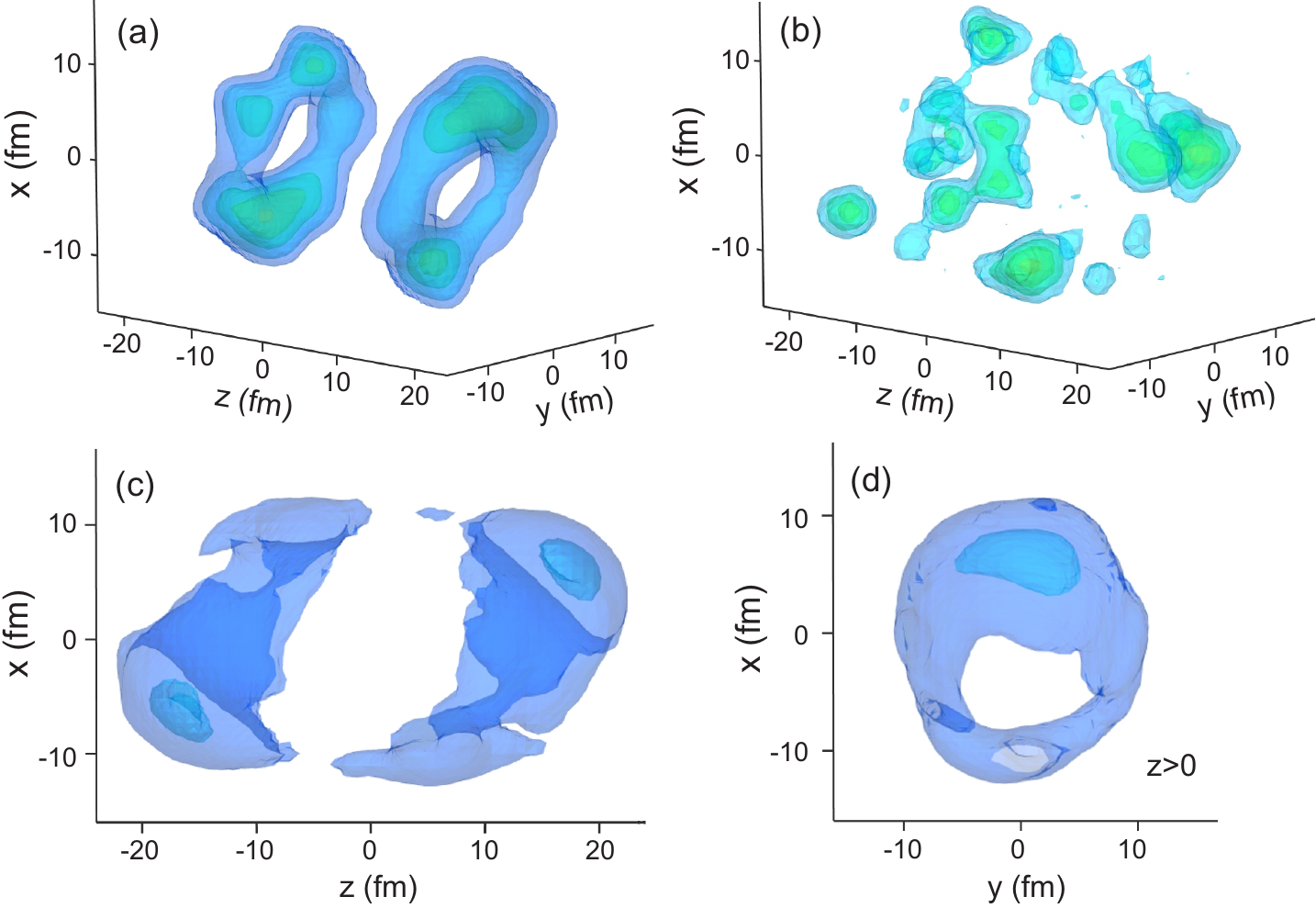}
\caption{Isosurfaces of nuclear density at $t=200 \, \text{fm}/c$ in Au + Au collisions at $60 \, \text{MeV/u}$ and $b= 2\, \text{fm}$ as predicted by the pBUU model [panel (a)] and the BM model [panels (b)-(d)]. The surfaces, following the order of shifting palette, are displayed for $\rho/\rho_0 = 0.1, 0.2, 0.4$ and~0.6, except for panel (b) where the lowest-$\rho$ surface is omitted.  Panel (b) displays density surfaces for a single collision event and panels (c) and (d) illustrate, respectively, side and front ($z>0$ only) views of surfaces for densities averaged over 100 events.  For more explanation see text.}
\label{fig:3D}
\end{figure}

Next we turn to results of the BM model and similarities and differences with pBUU. At early times, until $t \sim 140 \, \text{fm}/c$, the density evolves in BM in a similar manner as in pBUU.  As the reaction progresses, the comparison with pBUU depends on the incident energy.  At higher incident energies, growth of non-uniformities due to fluctuations competes with a more rapid expansion. At late times there, the BM density evolves towards smoother shapes, more like in pBUU simulations~\cite{lin_one-body_2019}.

Fig.~\ref{fig:3D} shows density isosurfaces at $t=200 \, \text{fm}/c$ in Au + Au collisions at $60 \, \text{MeV/u}$ and $b=2 \, \text{fm}$. In panel~(a) the density distribution in the pair of rings, predicted to form in pBUU (see Fig.~\ref{fig:contour}), is shown in 3D. In contrast, a single BM event at the same time and beam energy, in panel~(b), yields separated fragments, each moving at somewhat different velocity, reflecting impact of the fluctuations inserted with a physical justification.  Averaged over 100 events these structures consolidate in a compact form, shown in panels~(c) and~(d), analogous to pBUU.

\begin{figure}
\centering
\includegraphics[width=.65\linewidth]{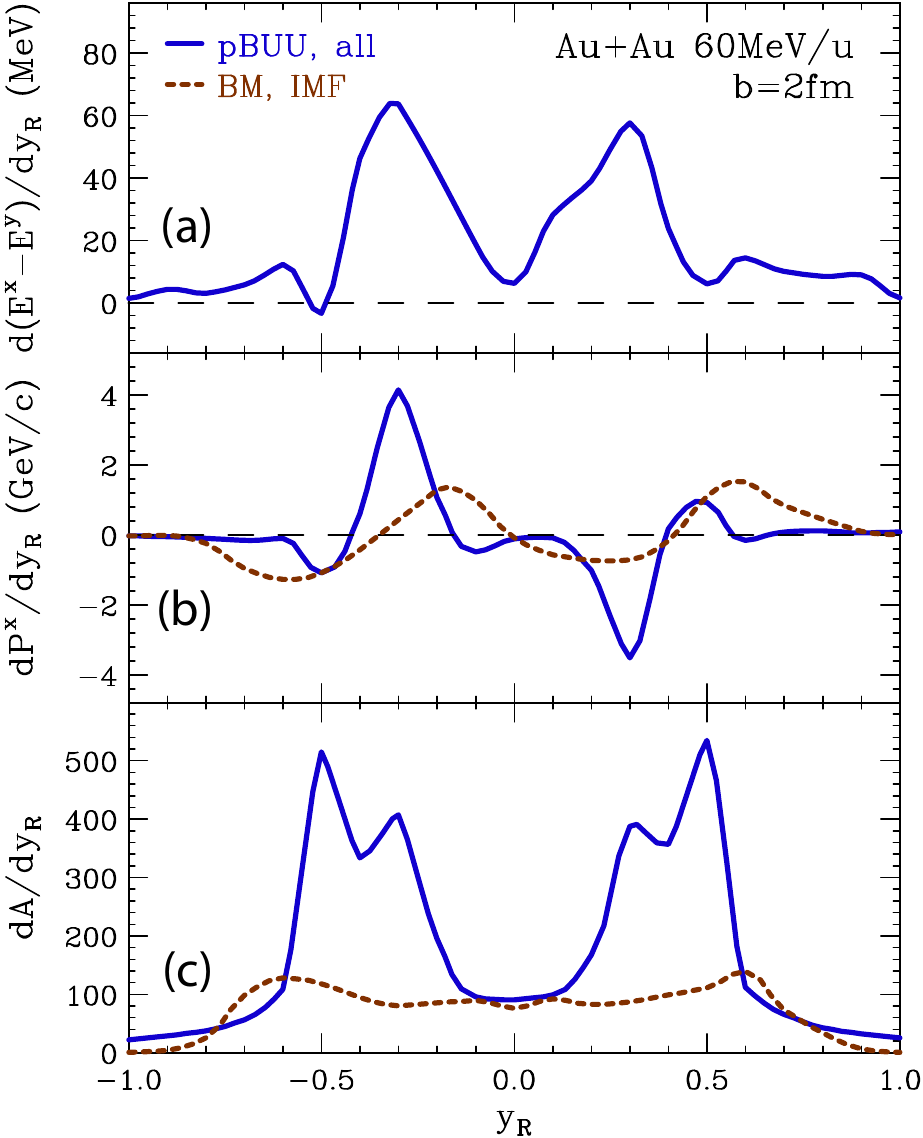}
\caption{Distribution of the asymmetry in the net transverse E$^x$-E$^y$ (in-out reaction plane) energy [panel (a)], the in-plane net momentum P$^x$ [panel (b)] and of the baryon number $A$ [panel (c)] as a function of rapidity $y_R$, normalized to beam, predicted for all particles in pBUU (solid lines), and for IMFs in BM (dashed). In the BM model, the transverse energy, averaged over 100 events, was too noisy to allow a meaningful inspection of the differential energy asymmetry in the ring region and that asymmetry was dropped for BM from panel (a).}
\label{fig:dNpx}
\end{figure}

Relevance of predictions of any transient shapes forming in HI due to instabilities depends on the ability to demonstrate their presence experimentally.  Due to the expansion involved in the ring formation, we observe some mapping of the ring pattern from the configuration space onto the velocity space. In measurements, patterns correlated with the reaction plane in the velocity space are quantified with transverse energy, momentum and direction vector moments studied vs rapidity~\cite{reisdorf_collective_1997,Danielewicz:2002pu,andronic_excitation_2005,le_fevre_constraining_2016}, such as next illustrated in Fig.~\ref{fig:dNpx}.  To arrive at the figure we weight local velocities at the late stages of collisions ($t \gtrsim 220 \, \text{fm}/c$), with local baryon densities.  The plotted distributions freeze eventually, i.e., become independent of the evaluation time $t$.

In our current simulations we have two rings, tilted relative to the beam axis, that are moving forward and backward and expanding.  In Fig.~\ref{fig:dNpx}(c), the leading edges of the rings in space end up as leading edges in the rapidity distribution $\text{d}P^x/\text{d}y_R$.  As the rings expand in their plane, the leading part of each ring yields a transverse momentum in one direction and a trailing edge in the opposite, hence a changing sign for $\text{d} P_x/\text{d}y_R$~\cite{danielewicz_transverse_1985} when rapidity spans the ring extension in $y_R$ in Fig.~\ref{fig:dNpx}(b). Overall, the transverse momentum in the reaction plane executes two oscillations along rapidity, contrasting with a single oscillation observed at both higher and lower energies~\cite{magestro_disappearance_2000,danielewicz_collective_1988}.  In the energy asymmetry, a notch appears for each ring, Fig.~\ref{fig:dNpx}(a), consistent with a transverse expansion of the rings.

The proposed experiment is to measure the changes in the rapidity distributions of observables depicted in Fig.~\ref{fig:dNpx}.   Detection of the predicted change of the sign of the transverse in-plane momentum can be challenging on the analysis side due to difficulties in determining the reaction plane.  However, flow characteristics can be determined without an explicit reaction-plane determination, specifically by examining scalar products of transverse momenta, and more generally transverse tensor convolutions, between different rapidities, and by looking for dominant factorizable contributions~\cite{danielewicz_collective_1988,cms_collaboration_principal-component_2017}. Following the relatively low transverse collective velocities at the onset of collective expansion, the fragments from the forward ring in the simulations are going to populate low angles of $5^\circ\text{--}7^\circ$ in the laboratory frame, so the experiment would benefit from forward detectors for registering these fragments.

We encourage an experimental search for qualitative fingerprints of the physical picture proposed in this work, and not to expect quantitative details of one model or the other to be met.  Quantitative predictions are a challenge around threshold.  Physically the fluctuations are expected to be there in the dynamics and leave imprints in the final state and these are modeled in BM.  On the other hand, aspects of mean field and fundamental collisions in the pBUU framework have more physical, better tested features.  The beam energy window, where the rings are predicted, includes $60 \, \text{MeV/u}$ for Au + Au collision in both models, but it is somewhat narrower in BM.

In summary, we find, for the first time that a combination of stalled equilibration, the spinodal instability at low densities in the matter and the rise of collective expansion with incident energy in central collisions leads to formation of transient ring-like structures both in the projectile and target residual systems. The joint application of both models addresses the question of the role of statistical fluctuations, absent in pBUU transport model and present in BM, that have potential to erase impacts of the ring formation. In either model, the transient structures, predicted in the configuration space, correlate with patterns in the velocity space thus allowing their experimental confirmation.  Our simulations provide specific guidance for the experimental strategy to investigate IMFs and thus demonstrate the presence of the liquid-gas transition and the spinodal instabilities experimentally.

\section*{Acknowledgements}

The authors benefited from discussions with Scott Pratt.
This work was supported by the U.S.\ Department of Energy Office of Science under Grant {DE}-{SC}0019209.

\bibliographystyle{elsarticle-num}

\bibliography{rings}

\end{document}